\documentclass[reprint,amsmath,amssymb,aps,pra,showpacs]{revtex4-1}
\usepackage{color}
\usepackage{graphicx}	
\usepackage{dcolumn}
\usepackage{bm}			
\usepackage{amsfonts}
\usepackage{dsfont}

\begin{document}

\preprint{APS/123-QED}

\title{Polynomial description of inhomogeneous topological superconducting wires}
 \author{Marcos P{\'e}rez$^{1,}$} \email{marcos.perez@ufrgs.br} 
 \author{Gerardo Mart\'{\i}nez$^{1,2}$}
  \affiliation{$^1$Instituto de F\'{\i}sica, UFRGS, 91501-970 Porto Alegre-RS, Brazil \\
               $^2$Instituto de F\'{\i}sica Te{\'o}rica UAM/CSIC, Madrid, Spain}
 
 \date{\today}

\begin{abstract}
We present universal features of the topological invariant of {\it p}-wave superconducting wires after the inclusion of spatial inhomogeneities. Three classes of distributed potentials are studied, a single-impurity, a commensurate and an incommensurate model using periodic site modulations. An analytical polynomial description is achieved by splitting the topological invariant into two parts, one dependent on the chemical potential and the other not. For the homogeneous case, an elliptical region is found where the topological invariant oscillates. The zeros of these oscillations occur at points where the fermion parity switches for finite wires. The increase of these oscillations with the inhomogeneity strength leads to new non-topological phases. We characterize these new phases according to each class of spatial distributions. Such phases could also be observed in the XY model.
\end{abstract}


\maketitle

%%%%%%%%%%%%%%%%%%%%%%%%%%%%%
%%%%%%%%%%%%%%%%%%%%%%%%%%%%%
\section{Introduction} 
%%%%%%%%%%%%%%%%%%%%%%%%%%%%%
%%%%%%%%%%%%%%%%%%%%%%%%%%%%%

Topological phases arising in superconducting materials have become a major research topic in condensed matter physics in the recent years.\ \cite{Alicea2012, Beenakker2013, Elliott2015} Indeed, the search for Majorana zero modes observation in lab has already begun.\ \cite{Mourik2012, Deng2012, Das2012} The potential application of these quantum phases is sustained by their robustness against a weak disorder, suggesting its realization in fault tolerant quantum computing.\ \cite{HasanKane2010, SDSarma2015} Hence, it is important to survey the effect of disorder in such topological phases.

A paradigmatic model in this context is the Kitaev chain, \cite{Kitaev2001} a lattice version of a spinless {\it p}-wave superconducting wire, whose topological phase has a bulk gap protected by time-reversal and particle-hole symmetries. The properties of this model, including spatial modulations, have been broadly and intensely studied.\ \cite{Lang2012, DeGottardi2013PRL, DeGottardi2013PRB, Wakatsuki2014, Hegde2015, Hegde2016, Zeng2016, YiGao2015, 2017arXiv170503783L}  Moreover, complex cases including symmetry breaking interactions, like spin-orbit or quartic fermionic terms, have been proposed to study recent experiments.\ \citep{Alicea2010, Ortiz2014, Adagideli2014, Queiroz2016} Yet, the key ingredients of the Kitaev free model can deal with the essential aspects of the topology with or without interactions, since all strongly interacting topological phases can be represented by noninteracting systems.\ \cite{Fidkowski2010, Tang2012, Gergs2016}

There are some previous works dedicated to study the case of modulated chemical potentials,\ \cite{Lang2012, DeGottardi2013PRL, DeGottardi2013PRB, Wakatsuki2014, Hegde2015, Hegde2016, Zeng2016} while other works deal with a modulated hopping.\ \cite{YiGao2015, 2017arXiv170503783L} They found, exploiting among other tools a scaled chemical potential,\ \cite{DeGottardi2013PRL, DeGottardi2013PRB} that the  spatial distributions strongly modifies the original topological phase diagram of the Kitaev model. Here, we are able to explain the origin of those changes by introducing a topological invariant which is characterized by an oscillatory polynomial delimited by two values. In particular, we can predict how new non-topological compact phases emerge with the inhomogeneities and characterize all changes induced on the  phase diagram. For example, in the homogeneous case the zeros of the oscillatory function of our approach are exactly at the positions where the fermion parity of the ground state for finite wires switches, positions that were previously obtained by other approaches. \cite{Hegde2015, Hegde2016}  We therefore propose that our method can be further used to describe the fermion parity switches in the general inhomogeneous case. We also propose to use this method to detect and map new paramagnetic phases in the ferromagnetic region of the phase diagram of the dual XY model, when the transverse field or any other exchange coupling constant are equally modulated. The latter is possible because the exact map of the Kitaev chain into the XY model through a non-local Jordan-Wigner transformation.\ \cite{LIEB1961407, Bunder1999}

The paper is organized as follows: In Section II, we introduce the model, present the different site distributions that we use and indicate the symmetry properties that classify it in  the BDI symmetry class. The $\mathds{Z}$-type topological invariant is constructed using the Zak phase. In Section III, we present our approach of separating the structural function of the topological invariant into two parts, one involving the chemical potential and the other not. In Section IV, we describe the polynomial structure of the topological invariant for the homogeneous case, the modulated chemical potential, and the modulated hopping amplitude, respectively. In Section V we summarize our results. Some appendices compile useful information about the polynomial description.

%%%%%%%%%%%%%%%%%%%%%%%%%%%
%%%%%%%%%%%%%%%%%%%%%%%%%%%
\section{The Model}
%%%%%%%%%%%%%%%%%%%%%%%%%%%
%%%%%%%%%%%%%%%%%%%%%%%%%%%

Consider a periodically modulated Kitaev model, with site-dependent chemical potentials and hopping amplitudes through an enlarged unit-cell -- of size $q$ -- repeated along the chain, while using periodic boundary conditions. Generically, for a system with $N$ sites we can write such model Hamiltonian as ${\cal H} = \sum_{\ell=1}^{N/q} \, {\cal H}_{\ell}$, where

\begin{eqnarray} \!\!\!
{\cal H}_{\ell} &=& \sum_{s=1}^{q} \mu_s^{} c_{s,\ell}^{\dag} c^{}_{s,\ell} + \nonumber \\ 
 && \sum_{s=1}^{q-1} \left( -t_s^{} c_{s,\ell}^{\dag} c^{}_{s+1,\ell} + \Delta_s^{} c^{}_{s,\ell} c^{}_{s+1,\ell} \right) + \nonumber \\ 
 && \left(-t_q^{} c_{q,\ell}^{\dag} c_{1,\ell+1}^{} +  \Delta_q^{} c^{}_{q,\ell} c^{}_{1,\ell+1} \right) + \mbox{H.c.}  \label{Hamilt}
\end{eqnarray}

\noindent
 ${\cal H}_{\ell}$ is a tight-binding model of spinless fermions with a site-dependent chemical potential ($\mu_s$), hopping amplitude ($t_s$), and {\it p\/}-wave superconducting pairing  ($\Delta_s$). For the site modulations we use either $\mu_s = \mu\,(1+\lambda w_s)$ for the chemical potential or $t_s = t\,(1+\lambda w_s)$ for the hopping amplitude, while $\Delta_s=\Delta$ is kept real. The additional parameter $\lambda$ provides the strength of the inhomogeneity and $w_s$ are the spatial distributions, which are taken as

\begin{equation}
w_s = \left\{
	\begin{array}{clc}
		\delta_{s,1}			&,\quad \mbox{a single-defect} 	& \mbox{(S)} \\
		\cos(2\pi s /q) 		&,\quad \mbox{commensurate} 	& \mbox{(C)} \\
		\cos(2\pi s\, \beta)	&,\quad \mbox{incommensurate} 	& \mbox{(I)}
	\end{array}
		\right. \label{3dist}
\end{equation}

\noindent
where $\beta= (\sqrt{5}+1)/2$ is the golden ratio. The first of these distributions (S) needs no justification, while the others two (C,I) are in the class of Aubry-Andr\'e or Harper potentials, useful for study the interrelation of disorder and superconductivity.\ \cite{AubryAndre1980, Cai2013, Cai2014, Zeng2016}

Upon using the Fourier transformation in each unit cell, $c_{s,\ell}^{\dag} = \sqrt{q/N}\sum_k c_{s,k}^{\dag} e^{ikq \ell}$, with $k\in (-\pi/q,\pi/q]$, the reduced Brillouin Zone, and through the well-known Bogoliubov-de Gennes (BdG) transformation,\ \cite{de1999superconductivity} $\Psi_k = ( c_{1,k}^{\dag},\dots, c_{q,k}^{\dag}\,,c_{1,-k}^{},\ldots, c_{q,-k}^{})^T$, we can write down the Hamiltonian in momentum space as

\begin{equation}
{\cal H} = \frac{1}{2} \sum_k \Psi_k^{\dag} \widehat{H}_k^{} \Psi_k^{} =  
	\frac{1}{2} \sum_k 
	\Psi_k^{\dag} 
		\begin{pmatrix} 
			\hat{V}_k^{} & \hat{M}_k^{} \\ 
			\hat{M}_k^{\dag} & -\hat{V}_{-k}^{T} 
		\end{pmatrix} 
	\Psi_k^{} \,, \label{eq:Hk}
\end{equation}

\noindent
where $\hat{V}_k$ and $\hat{M}_k$ are $q\times q$ matrices whose non-zero elements are: $V_k^{s,s}=\mu_s$ for $s=1,\dots,q$\,; $V_k^{s,s+1} = V_k^{s+1,s}=-t_s$ and $M_k^{s,s+1}=-M_k^{s+1,s}=-\Delta$ for $s=1,\dots,q-1$\,;  while $V_k^{q,1}=(V_k^{1,q})^* = -t_q e^{-ikq}$ and  $M_k^{q,1}=(-M_k^{1,q})^*=-\Delta e^{-ikq}$ (see also Gao {\it et al.} \cite{YiGao2015}).

As long as the values of $\mu_s$, $t_s$ and $\Delta$ are real, this model has a time-reversal symmetry $\mathcal{T=K}$ ($\mathcal{K}$ takes the complex conjugate) that satisfies $\mathcal{T}\widehat{H}_k\mathcal{T}^{-1}=\widehat{H}_{-k}$, while the BdG transformation evinces a particle-hole symmetry of the model, characterized by  $\mathcal{P}=\tau_x\,\mathcal{K}$ (where $\tau_x$ is the Pauli matrix acting on the particle-hole space) that satisfies $\mathcal{P}\widehat{H}_k\mathcal{P}^{-1}=-\widehat{H}_{-k}$. With these two  operators, we can build a chiral operator $\mathcal{C}=\mathcal{TP}=\tau_x$ that satisfies $\mathcal{C}\widehat{H}_k\,\mathcal{C}^{-1}=-\widehat{H}_k$. For the operators described above, we have $\mathcal{T}^2 = \mathcal{P}^2=\mathcal{C}^2=1$, classifying this 1D system into the BDI symmetry class,\ \cite{AltlandZirnbauer1997, Schnyder2008} whose topological invariant is characterized by a $\mathds{Z}$-index.

The topological invariant can be built from the Zak phase,\ \cite{Zak1989, STewari2012, Xiao2010} which is the Berry phase for periodic fermion systems. To implement it, we first rotate the chiral Hamiltonian (\ref{eq:Hk}) into a purely off-diagonal form

\begin{equation}
\Omega\, \widehat{H}_k \, \Omega^{\dag} = 
	\begin{pmatrix}
		0 & \hat{A}_k \\ 
		\hat{A}^T_{-k}& 0
	\end{pmatrix},\quad 
		\hat{A}_k = \hat{V}_k + \hat{M}_k\,, 
\label{off-diag}
\end{equation}

\noindent
where $\Omega= e^{-i\pi\tau_y/4}$ is the unitary transformation and $\tau_y$ is the Pauli matrix acting on the particle-hole space. The topological index is thus the winding number of the eigenstates in the reduced Brillouin Zone, which for this kind of off-diagonal matrix is given by \cite{STewari2012}

\begin{equation}
\mathcal{W}= -\frac{i}{\pi}\int_{k=0}^{\pi/q}\frac{dz_k}{z_k}\,, \,\,\, 
	\mbox{where} \quad z_k = 
	\frac{\mbox{Det}(\hat{A}_k)}{|\mbox{Det}(\hat{A}_k)|}\,.
\end{equation}

\noindent
In this approach, the winding number ($\mathcal{W} \in \mathds{Z}$) can be evaluated through the sign of the function $\mbox{Det}(\hat{A}_{k})$ at the particle-hole symmetric $k$ points,  $0$ and $\pi/q$

\begin{equation}
\mathcal{W} = \frac{1}{2} \left[ \, sgn\{\mbox{Det}(\hat{A}_{\pi/q})\}- sgn\{\mbox{Det}(\hat{A}_0)\} \, \right].
\label{eq:windingNum}
\end{equation}

\noindent
Consequently, for the Hamiltonian in Eqs.\ (\ref{eq:Hk}) and (\ref{off-diag}) the winding number $\mathcal{W}$ can only be $\pm 1$ or 0 (Topological/Non-Topological, respectively). 
We further notice that $\hat{A}_k= (\hat{A}_{-k})^*$, thereby, $\mbox{Det}(\hat{A}_k)$ is a real (polynomial) function at the extreme points, $\kappa=0$ and $\pi/q$, providing a well-defined mathematical expression (\ref{eq:windingNum}), from which a $\mu$-$\Delta$ phase diagram can be set up.

\begin{figure}[t] \centering
\includegraphics[width=7.90cm]{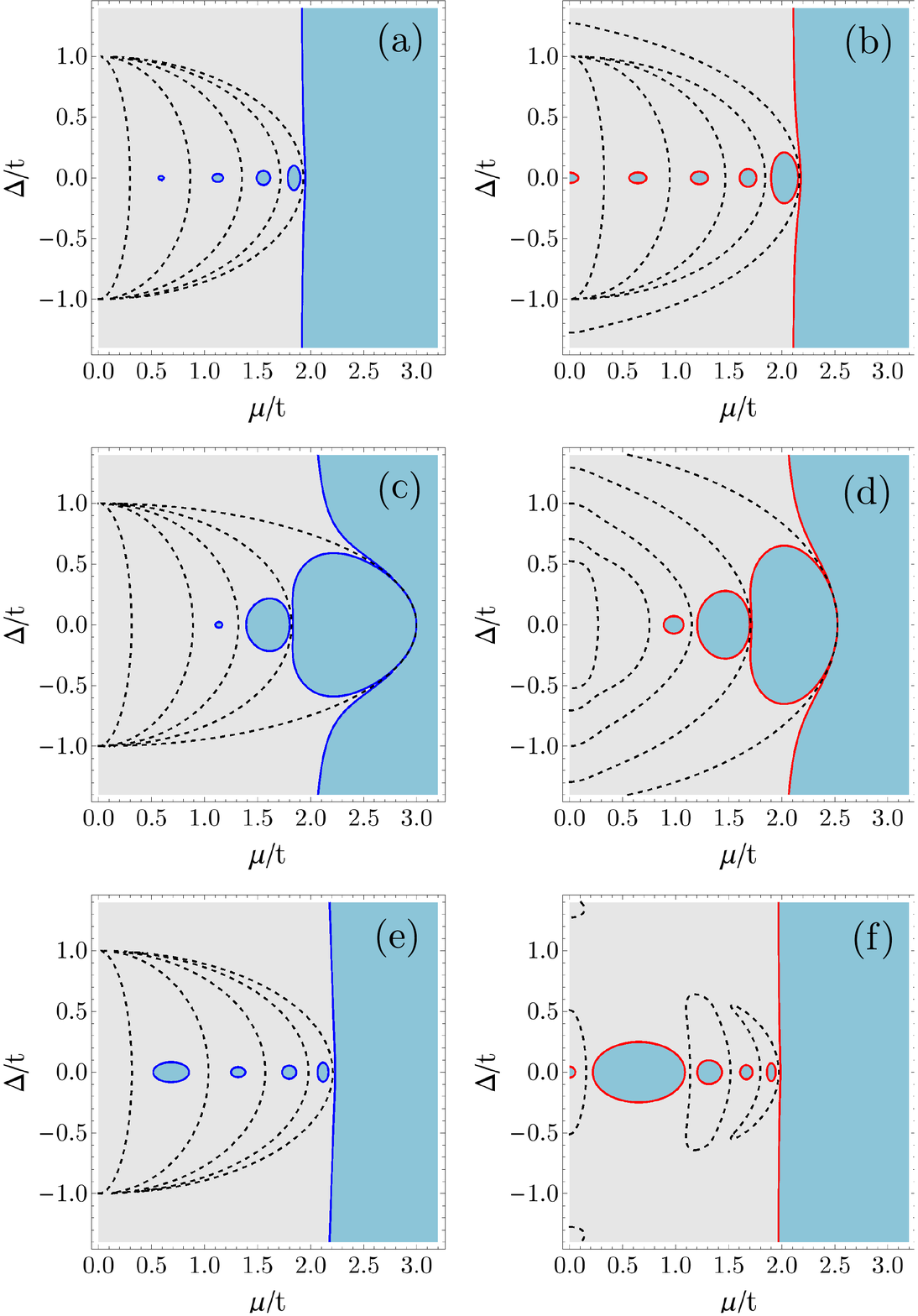}
\caption{Topological phase diagrams  for modulated chemical potentials (left panels) and modulated hopping terms (right panels) in the (a,b) single-defect model, (c,d) commensurate and (e,f) incommensurate potentials. Topological ($T$) phases are the regions in light gray and Non-Topological ($NT$) phases are the zones in blue. Dark blue (left panels) and red (right panels) solid lines are the {\it loci} of the $T$-$NT$ phase transitions, for which $\mbox{Det}(\hat{A}_\kappa)=0$. Dashed lines depicting ellipses are zeros of the oscillating function $U(\mu,\Delta,q,\lambda)$ given below. All diagrams are symmetric by changing $\mu \leftrightarrow -\mu$ and $\Delta \leftrightarrow -\Delta$. In all cases, we used $q=10$ and $\lambda=0.5$ in the enlarged unit-cell approach. Emerging features are the non-topological regions (``bubbles'') around the anisotropy line ($\Delta/t=0$), and the bending of the Ising transition lines at $\mu\approx 2t$, notably for (c) and (d) cases. Case (f) will be discussed separately.}
\label{fig:domehomo}
\end{figure}

Some selected results for the spatial distributions (S,C,I) can be viewed in Fig.\ \ref{fig:domehomo}.\ Apart from changes of the Ising Topological/Non-Topological ($T$-$NT$) transition lines at $|\mu|=2t$, we observe in all cases the emergence of non-topological compact domains (``bubbles'') around $\Delta/t=0$. The number and shape of these bubbles depend on the cell size $q$, on the inhomogeneity strength $\lambda$, as well as on the spatial distribution. They look rather different when spatial modulations are applied either to the chemical potential or to the hopping amplitude, as can be seen from Fig.\ \ref{fig:domehomo} (left and right panels, respectively).

%%%%%%%%%%%%%%%%%%%%%%%%%%%%%%%%%%%%%%%%%
%%%%%%%%%%%%%%%%%%%%%%%%%%%%%%%%%%%%%%%%%
\section{The Topological Invariant}
%%%%%%%%%%%%%%%%%%%%%%%%%%%%%%%%%%%%%%%%%
%%%%%%%%%%%%%%%%%%%%%%%%%%%%%%%%%%%%%%%%%

For the origin of the emerging bubbles in the topological regions of the phase diagram for the inhomogeneous cases, we  consider some analytical aspects of the topological invariant. First, we rewrite the (real) function $\mbox{Det}(\hat{A}_{\kappa})$, for $\kappa=0$ and $\pi/q$, as

\begin{equation}
\mbox{Det}(\hat{A}_{\kappa}) = U(\mu,\Delta,q,\lambda)  + \Lambda_{\kappa}(\Delta,q,\lambda) ,
\label{DetAk}
\end{equation}

\noindent
where $U$ is a $\kappa$-independent polynomial function that carries all the dependence on $\mu$, and $\Lambda_{\kappa}$ is the difference, in our $(q,\lambda)$ scheme. According to Eq.\  (\ref{eq:windingNum}), we observe that the system is topological ($\mathcal{W}\neq 0$) when $U$ is within the region delimited by both $\Lambda_{\kappa}$ ({\it viz}.: $\Lambda_0 < U < \Lambda_{\pi/q}$) and non-topological ($\mathcal{W}=0$) otherwise. For cases where $\mbox{Det}(\hat{A}_\kappa)=0$ Eq.\ (\ref{eq:windingNum}) is undefined, though these points determine the {\it loci} of the $T$-$NT$ transitions in the phase diagram, namely, the closure of the bulk gap, as is inferred from Eq.\ (\ref{off-diag}). We have selected two featuring examples to display quantitatively this situation.

In Fig.\ \ref{fig:TPD} we plotted the homogeneous case (left panels) and the inhomogeneous single-defect model applied to the chemical potential (right panels), respectively. In the homogeneous case, we see from Fig.\ \ref{fig:TPD}(c) that the function $U$ oscillates between the two $\Lambda_{\kappa}$ functions, for $|\mu|\leqslant 2t$, defining thus a topological phase in that region. Whereas in the single-defect case of Fig.\ \ref{fig:TPD}(d), the function $U$ leaves these two limits because its oscillations are now enhanced by the presence of modulations. Beyond the topological phase, in the $\mu/t$ axis, the function $U$ changes character and becomes unbounded in all cases. These trending features deserve further analytical  explanations.

%%%%%%%%%%%%%%%%%%%%%%%%%%%%%%%%%%%%%
%%%%%%%%%%%%%%%%%%%%%%%%%%%%%%%%%%%%%
\section{Polynomial structure}
%%%%%%%%%%%%%%%%%%%%%%%%%%%%%%%%%%%%%
%%%%%%%%%%%%%%%%%%%%%%%%%%%%%%%%%%%%%

\subsection{The homogeneous case}

After detailed algebraic and numerical manipulations we found, in the homogeneous case ($\lambda=0$), that the function $U$ can be written as

\begin{equation}
 U(\mu,\Delta,q,0) = \left(\!\sqrt{1-(\Delta/t)^2}\,\right)^q \, \overline{U}_H(\bar{\mu},q) ,
 \label{functionU}
\end{equation}

\noindent 
where $\bar{\mu} = \mu/\sqrt{1-(\Delta/t)^2}$ is a scaled chemical potential, identical to the one found in DeGottardi {\it et al.}, \citep{DeGottardi2013PRL, DeGottardi2013PRB} while $\overline{U}_H(\bar{\mu},q)$ is described by a polynomial in $\bar{\mu}$ of degree $q$, restricted to integers $q\geqslant 2$

\begin{equation}
 \overline{U}_H(\bar{\mu},q) = \sum_{n=1}^q a^q_{n} \, \bar{\mu}^n ,
 \label{eq:UH}
\end{equation}

\noindent
whose coefficients $a^{q}_{n}\in \mathds{Z}$ can be obtained through the following recurrence formula: $a^q_q=1$ for all $q$, $a^q_1=\mp q$ (alternating) for $q$ odd and $a^{q}_{1}=0$ for $q$ even, while for $1<n<q$ we have $a^{q}_{n}=a^{q-1}_{n-1}-a^{q-2}_{n}$ for $q-n$ even and $a^q_n=0$ for $q-n$ odd. More details can be found in the Appendix A.

As for the delimiting functions $\Lambda_{\kappa}$ in Eq.\ (\ref{DetAk}), we found a useful relation in the form of a sum rule

\begin{equation}
 \Lambda_{\pi/q} + \Lambda_0 = \left(\!\sqrt{1-(\Delta/t)^2}\,\right)^q s^{1/2}\,(1+s)^2 ,
 \label{sumrule}
\end{equation}

\begin{figure}[t] \centering 
\includegraphics[width=7.80cm,angle=-90]{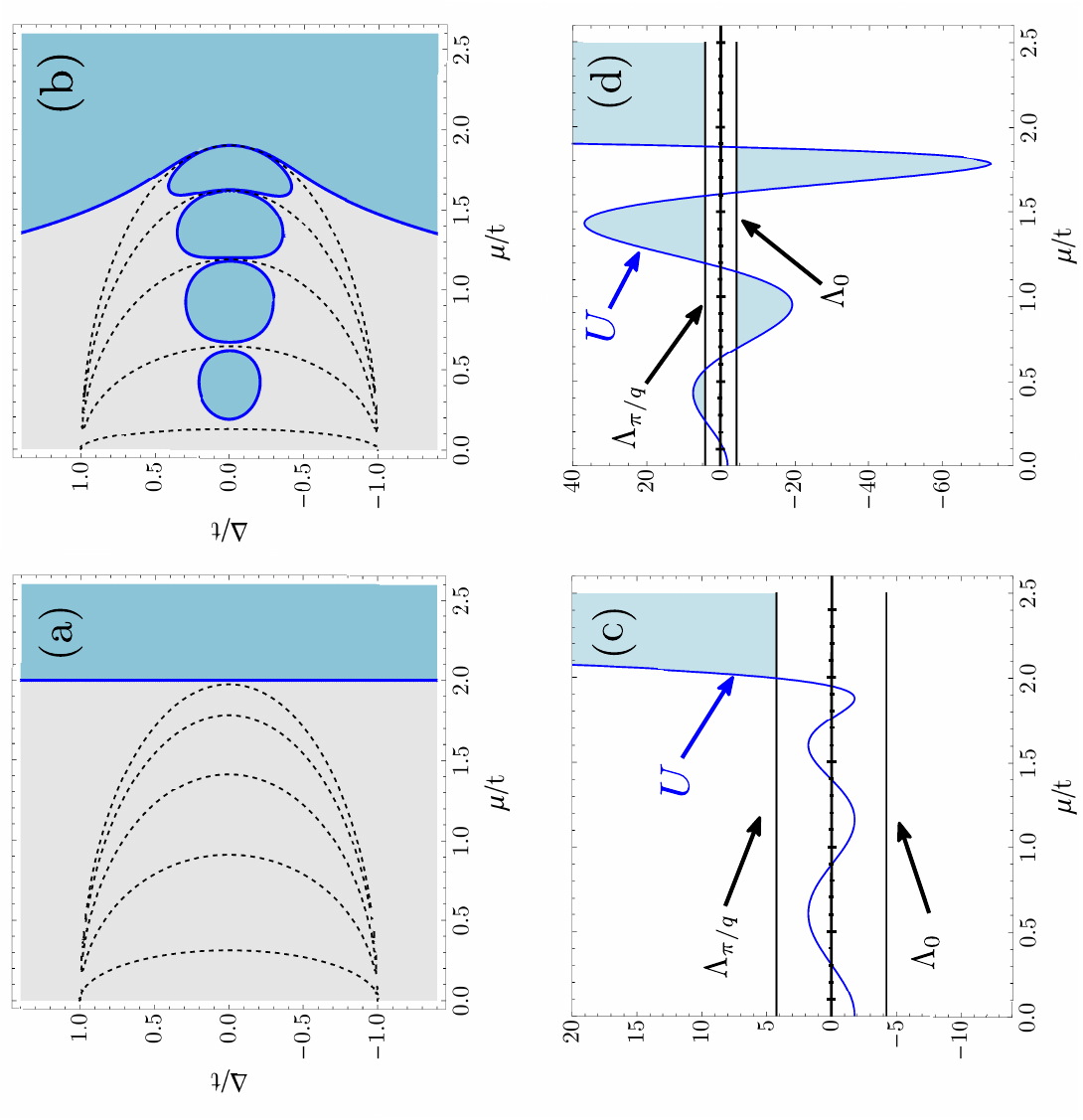}
\caption{Comparison of the homogeneous case (left panels, $\lambda=0$) and the inhomogeneous single-defect case (S) applied to the chemical potential (right panels, $\lambda=10$). Topological regions are in gray, non-topological in blue. Dashed lines in phase diagrams (a) and (b) are zeros of the oscillating function $U$, whose outermost ones are very close to the ellipse $\bar{\mu}=2t$, beyond which $U$ is overdamped. Function $U$ (blue curves) in (c) and (d) oscillates with $q$ zeros until it crosses the Ising transition line close to $\mu/t=2$. The main difference is that in (c), the homogeneous case, $U$ is delimited by the two $\Lambda_{\kappa}$  functions, while in (d) $U$ goes beyond these two limits. This is the origin of the bubbles observed in all inhomogeneous cases. These bubbles have alternating sign, corresponding to the wavefunction fermion parity. We have used $\Delta/t=0.2$ in panels (c) and (d). For drawing purposes, we also shifted the vertical axis to get $U=0$ as the mean of both $\Lambda_{\kappa}$ functions. }
\label{fig:TPD}
\end{figure}

\noindent
where $s=(-1)^q$ and the factor $(1+s)^2$ dictates that this sum is zero for $q$ odd, while for $q$ even the sum rule can be used to shift the zero. Using (\ref{sumrule}), we thus need to deal with only one of these polynomials. In this work, we study $\Lambda_0=\Lambda_0 (\Delta,q)$ which is given, in the homogeneous case, by an even polynomial in $\Delta$

\begin{equation}
 \Lambda_0(\Delta,q) = \sum_{n=0}^{q} b_n^q \, \Delta^n .
 \label{lambdazero}
\end{equation}

\noindent
Some coefficients $b_n^q$ are given in the Appendix A. No simple recurrence formula for $b_n^q$ has been found so far.

An interesting point is that the alternating integer polynomial $\overline{U}_H(\bar{\mu},q)$ in (\ref{eq:UH}) has more than one root (in fact, $q$ real roots), that yield an oscillatory behavior for $U$  within the dome $\bar{\mu}=2t$, that is, inside the ellipse $(\mu/2t)^2 + (\Delta/t)^2 = 1$ (see Fig.\ \ref{fig:TPD}(a)). Outside from this elliptical dome $U$ has no more oscillations, as $\overline{U}_H$ turns from an alternating to a positive  polynomial in $\mu$ in that case. Such analytical behavior is a critical combination of the prefactor $(\!\sqrt{1-(\Delta/t)^2}\,)^q$ and the scaled  $\bar{\mu} = \mu/\sqrt{1-(\Delta/t)^2}$ that appear in Eq. (\ref{functionU}). This result is consistent with those seen in Hegde {\it et al.}, \cite{Hegde2015, Hegde2016} about the positions of fermion parity switches for finite wires.

%%%%%%%%%%%%%%%%%%%%%%%%%%%%%%%%%%%%%%%%%
\subsection{Modulated chemical potential}
%%%%%%%%%%%%%%%%%%%%%%%%%%%%%%%%%%%%%%%%%

When site modulations are applied to the chemical potential the sum rule (\ref{sumrule}) still applies, therefore the polynomial expansion of $\Lambda_0(\Delta,q)$ is the same as in (\ref{lambdazero}), while the function $U$ can be written now as

\begin{equation}
 U(\mu,\Delta,q,\lambda) = \left(\!\sqrt{1-(\Delta/t)^2}\,\right)^q \left( \, \overline{U}_H +  \overline{U}_\lambda \right) .
 \label{U_S}
\end{equation}

\noindent
For the (S) single-defect case, the extra inhomogeneous contribution is linear on $\lambda$, that is $\overline{U}_\lambda=\lambda\,\overline{U}_S(\bar{\mu},q)$, where $\overline{U}_S$ is a polynomial like in (\ref{eq:UH}), with the same recurrence formula for the coefficients $a^{q}_{n}$, except that $a^{q}_{1}=\mp 1$ (alternating) for $q$ odd and $a^{q}_{1}=0$ for $q$ even. See more details in the Appendix B.

For the commensurate (C) and the incommensurate (I) potentials applied to $\mu_s$, the extra contributions do not depend on $\Delta$ but are nonetheless nonlinearly dependent on $\lambda$. That is, we have integer polynomials $\overline{U}_\lambda$ in (\ref{U_S}) written as $\overline{U}_C(\bar{\mu},q,\lambda)$ and $\overline{U}_I(\bar{\mu},q,\lambda)$, respectively, for which we have not found a simple recurrence formula yet. See the examples, for $q=3$, in the Appendix B.

%%%%%%%%%%%%%%%%%%%%%%%%%%%%%%%%%%%%%%%%
\subsection{Modulated hopping amplitude}
%%%%%%%%%%%%%%%%%%%%%%%%%%%%%%%%%%%%%%%%

Significant differences already begin when spatial modulations are applied to $t_s$, the hopping amplitude. For the (S) single-defect case, for example, we have found a function $U$ described as
\begin{eqnarray} 
U(\mu,\Delta,q,\lambda) &=& \left(\!\sqrt{1-(\Delta/t)^2}\,\right)^q  \nonumber \\
&\times & \left( \overline{U}_H + \frac{\lambda(\lambda+2)}{[1-(\Delta/t)^2]} \, {\tilde{U}}_{S} \right) , 
\label{tildeU_s}
\end{eqnarray}

\noindent
Likewise, the sum rule for the $\Lambda_{\kappa}$ functions is now 
\begin{eqnarray}
\Lambda_{\pi/q} + \Lambda_0 & = & \left(\!\sqrt{1-(\Delta/t)^2}\,\right)^q s^{1/2}\,(1+s)^2 \nonumber  \\ 
&\times & \left( 1 + \frac{\lambda(\lambda+2)}{[1-(\Delta/t)^2](1+s)} \right) .
\label{Lambda}
\end{eqnarray}

\noindent
In (\ref{tildeU_s}) we have $\tilde{U}_S=\tilde{U}_S (\bar{\mu},q)$, which is a polynomial in $\bar{\mu}$ of degree $q\geqslant 3$, given by the same recurrence formula as in (\ref{eq:UH}), except that the largest-power non-zero coefficients are now given by $a^q_{q-2}=-1$ ({\it viz}.: $a^q_q=a^q_{q-1}=0$ for all $q$). More details are given in Appendix C. We notice in Eqs.\ (\ref{tildeU_s}) and (\ref{Lambda}) the extra terms in $\lambda(\lambda+2)$, which depend also on $\Delta$. This contribution expands the dome of the oscillations of the function $U$ beyond the limits $\Delta/t=\pm1$, as is indeed observed in Fig.\ \ref{fig:domehomo}(b).

For the commensurate (C) and the incommensurate (I) potentials applied to the hopping, on the other hand, a general expression for the function $U$ is

\begin{equation}
 U(\mu,\Delta,q,\lambda) = \left(\!\sqrt{1-(\Delta/t)^2}\right)^q   \overline{U}_H + \tilde{U}_{\lambda}.
 \label{tildeU_c}
\end{equation}

\noindent
The corresponding inhomogeneous $\tilde{U}_{\lambda}$ terms, namely, $\tilde{U}_{C}=\tilde{U}_{C}(\mu,\Delta,q,\lambda)$ or $\tilde{U}_{I}=\tilde{U}_{I}(\mu,\Delta,q,\lambda)$, respectively, are now polynomials in $\mu$ and $\Delta$, which cannot be simply factorized as in (\ref{U_S}) or in (\ref{tildeU_s}). The latter cases in (\ref{tildeU_c}) are examples where the above elliptical description does not apply anymore, as can be inferred from Figs.\ \ref{fig:domehomo}(d) and \ref{fig:domehomo}(f). The farthest case from an elliptical description is the latter one (f), for which the incommensurate (I) distribution provides a rather complex $U$ function. In these cases, we do not have integer polynomials anymore. Simple recurrence formulas for the terms $\tilde{U}_{C}$, $\tilde{U}_{I}$ or $\tilde{\Lambda}_{\kappa}$ in the modulated $t_s$ case were not yet found, although there seems to be a fractional/irrational polynomial expansion in the $\cos(2n\pi\beta)$ harmonics for some of them. Examples for $q=3$ and $5$, given in the Appendix C, suggest that the latter is true.

%%%%%%%%%%%%%%%%%%%%%%%%%%%%%%%%%%%%%
%%%%%%%%%%%%%%%%%%%%%%%%%%%%%%%%%%%%%
\section{Concluding Remarks}
%%%%%%%%%%%%%%%%%%%%%%%%%%%%%%%%%%%%%
%%%%%%%%%%%%%%%%%%%%%%%%%%%%%%%%%%%%%

In summary, we have studied the effects of spatial inhomogeneities in the 1D {\it p}-wave Kitaev model, for which we constructed an integer polynomial description for the topological invariant using the enlarged-unit cell approach. We applied the method to three different classes of spatial distributions, (S,C,I), although it can be easily expanded to more general cases, like those including longer-range hoppings and pairings. \cite{Viyuela2016, Alecce2017, Patrick2017}

A comparative study was made for modulated chemical potentials and modulated hopping amplitudes, finding very clear differentiations, as the examples from Fig.\ \ref{fig:domehomo} have demonstrated. The modulations applied to the chemical potential, for example, preserve the elliptical behavior mentioned above, while those applied to the hopping amplitude do not. This marked difference will establish a strong dichotomy for diagonal versus off-diagonal disordered systems.

The oscillatory behavior of the polynomial function $U$ and its delimiting functions take account of most of the topological features of the model, not only about the origin of emerging non-topological bubbles around the anisotropy line ($\Delta=0$) in the topological region of the phase diagram, but also of the exact positions of the ground-state fermion parity switches for homogeneous finite wires, which are given by the elliptical curves 
\[ \bar{\mu}=2t\cos \left(\frac{ \pi p }{ q+1 } \right) \] 

\noindent
where $p=1,2,\ldots,[q/2]$, with $[q/2]$ the integer part of $q/2$. See more details in Hegde {\it et al.} \cite{Hegde2015, Hegde2016}

The fact that our periodic boundary bulk results are consistent with those for open boundary systems, with edge Majorana fermions, obtained otherwise through the large size limit of the transfer matrix approach,\ \cite{Hegde2015, Hegde2016} is a consequence of the {\it bulk-edge correspondence}. Therefore, with this validity test we propose to use these results also for the inhomogeneous cases. The new zeros of the shifted function $U$, like those seen in the examples of Figs.\ \ref{fig:domehomo} and \ref{fig:TPD}, should be the positions of the new fermion parity switches in those cases.

Moreover, as a byproduct, since the Kitaev chain is exactly mapped into the XY model through a Jordan-Wigner transformation, \cite{LIEB1961407, Bunder1999} we  hope that other authors would be willing to observe paramagnetic bubbles in the ferromagnetic region of the inhomogeneous XY model, as well. What will remain to verify is whether the oscillations of the spin correlation function for the XY model would be at the same oscillatory region of the shifted function $U$, as it is in the homogeneous case.

%%%%%%%%%%%%%%%%%%%%%%%%%%%%
%%%%%%%%%%%%%%%%%%%%%%%%%%%%
\appendix
%%%%%%%%%%%%%%%%%%%%%%%%%%%%
%%%%%%%%%%%%%%%%%%%%%%%%%%%%

\section{The homogeneous case}

The homogeneous case ($\lambda=0$) is characterized by a function $U$ given by Eqs.\ (\ref{functionU}) and (\ref{eq:UH}). This factorization provides an integer polynomial $U_H$, whose coefficients are given by the recurrence formula: $a^q_q=1$ for all $q$, $a^q_1=\mp q$ (alternating) for $q$ odd and $a^{q}_{1}=0$ for $q$ even, while for $1<n<q$ we have $a^{q}_{n}=a^{q-1}_{n-1}-a^{q-2}_{n}$ for $q-n$ even and $a^q_n=0$ for $q-n$ odd. Some of these coefficients $a_n^q$ are given in Table \ref{table:Uhomo}. We notice that the nonzero coefficients of this polynomial expansion have alternating signs for each $q$. The oscillatory behavior of the function $U$ is due to this fact. It is easily seen that $U$, as a function of $\mu$ is symmetric for $q$ even and antisymmetric for $q$ odd, while it is always symmetric with respect to $\Delta$. For these appendices we shall take $t=1$.

\begin{table}[ht]
\centering
\begin{tabular}{|c|cccccccccc|}
\hline
 q &&&&&n&&&&&\\
 & 1 & 2 & 3 & 4 & 5 & 6 & 7 & 8 & 9 & 10 \\ 
\hline\hline
2 & 0 & 1 & 0 & 0 & 0 & 0 & 0 & 0 & 0 & 0 \\
3 & -3 & 0 & 1 & 0 & 0 & 0 & 0 & 0 & 0 & 0 \\
4 & 0 & -4 & 0 & 1 & 0 & 0 & 0 & 0 & 0 & 0 \\
5 & 5 & 0 & -5 & 0 & 1 & 0 & 0 & 0 & 0 & 0 \\
6 & 0 & 9 & 0 & -6 & 0 & 1 & 0 & 0 & 0 & 0 \\
7 & -7 & 0 & 14 & 0 & -7 & 0 & 1 & 0 & 0 & 0 \\
8 & 0 & -16 & 0 & 20 & 0 & -8 & 0 & 1 & 0 & 0 \\
9 & 9 & 0 & -30 & 0 & 27 & 0 & -9 & 0 & 1 & 0 \\
10 & 0 & 25 & 0 & -50 & 0 & 35 & 0 & -10 & 0 & 1 \\
\hline
\end{tabular}
\caption{Some coefficients $a_n^q$ of the integer polynomial $\overline{U}_H$ in the homogeneous case. See the rule $a^{q}_{n}=a^{q-1}_{n-1}-a^{q-2}_{n}$.}
\label{table:Uhomo}
\end{table}

\noindent
For the delimiting functions $\Lambda_{\kappa}$, on the other hand, a useful relation is given by the sum rule, Eq.\ (\ref{sumrule}), that allows to work with one of these polynomials. In this work, we deal with $\Lambda_0(\Delta,q)=\sum_{n=0}^{q} b_n^q \, \Delta^n$ which in this case is an even polynomial in $\Delta$ ($b_n^q=0$ for $n$ odd). Some of these coefficients are given in Table \ref{table:LambdaHomo}. These $b_n^q$ coefficients do not have a simple constructing rule, although we observe the cyclic pattern `2101' for $-b_0^q/2$, also $b_q^q=0$ for all $q$, $-b_{q-1}^q/2=q$ for $q$ odd, while for $q$ even we notice the sequence $-b_0^2/2=1\cdot2$, $-b_2^4/2=2\cdot4$, $-b_4^6/2=3\cdot6$, $-b_6^8/2=4\cdot8$, etc.

\begin{table}[ht]
\centering
\begin{tabular}{|c|ccccc|}
\hline
q &&&n&&\\
 & 0 & 2 & 4 & 6 & 8\\ 
\hline\hline
2 & 2 & 0 & 0 & 0 & 0 \\
3 & 1 & 3 & 0 & 0 & 0  \\
4 & 0 & 8 & 0 & 0 & 0  \\
5 & 1 & 10 & 5 & 0 & 0 \\
6 & 2 & 12 & 18 & 0 & 0 \\
7 & 1 & 21 & 35 & 7 & 0 \\
8 & 0 & 32 & 64 & 32 & 0 \\
9 & 1 & 36 & 126 & 84 & 9  \\
10 & 2 & 40 & 220 & 200 & 50 \\
\hline
\end{tabular}
\caption{Some coefficients $-b_n^q/2$ of the delimiting function $\Lambda_0(\Delta,q)$ in the homogeneous case.}
\label{table:LambdaHomo}
\end{table}

The oscillatory behavior of the function $U$ in the homogeneous case can therefore be summarized as follows: for $|\Delta/t|>1$, the prefactor $(\sqrt{1-(\Delta/t)^2})^q$ in Eq. (\ref{functionU}) is purely imaginary for $q$ odd and a negative number for $q$ even. This fact, together with the scaled chemical potential $\bar{\mu}=\mu/\sqrt{1-(\Delta/t)^2}$ in the polynomial expansion of $U_H(\bar{\mu},q)$, determine an overall effect on $U$ that change signs to all negative coefficients $a^q_n$, as seen from Table \ref{table:Uhomo}. These changes of sign cause $U$ to turn from an oscillatory to an overdamped polynomial with all coefficients positive. Consequently, for $|\Delta/t|>1$ the function $U$ does not oscillate in the homogeneous case. For $|\Delta/t|<1$, on the other hand, the function $U$ has an oscillatory part and an overdamped part. The oscillatory part of the shifted-$U$ is delimited by $\bar{\mu}=2t$. This constraint describes an ellipse in the parameter space, written as $(\mu/2t)^2+(\Delta/t)^2 = 1$, inside which the function $U$ oscillates, while outside it is just an overdamped function. This oscillatory region also explains the transition from the topological to the non-topological phase when $|\mu|=2t$, as the ellipse contains this limit when $\Delta/t=0$. Once $U$ is a positive integer polynomial it is unbounded and will then cross the region delimited by the two functions $\Lambda_\kappa$.

\section{Modulated chemical potential}

When spatial distributions are applied to the chemical potential, the coefficients of the polynomial function $U$ depend now on the inhomogeneity strength $\lambda$, while the delimiting functions $\Lambda_{\kappa}$ are not affected by $\lambda$, since the sum rule is the same as in the homogeneous description.

\subsubsection{Single-defect case}

When we apply the single-defect spatial distribution to the chemical potential, the function $U(\mu,\Delta,q,\lambda)$ can be factorized as in (\ref{U_S}), where $\overline{U}_H$ is the polynomial for the homogeneous case and $\overline{U}_S(\bar{\mu},q)$ is a polynomial in $\bar{\mu}$ of degree $q$, linearly dependent in $\lambda$. The coefficients $a^{q}_{n}$ of this polynomial have the same recurrence formula as in Eq.\ (\ref{eq:UH}), except that for $a^{q}_{1}=\mp 1$ (alternating) for $q$ odd and $a^{q}_{1}=0$ for $q$ even. Some of these coefficients are given in Table \ref{table:US-ChP}. Furthermore, the description of the function $\Lambda_{0}$ in this case is the same as the homogeneous case, as given in Table \ref{table:LambdaHomo}.

\begin{table}[t]
\centering
\begin{tabular}{|c|cccccccccc|}
\hline
 q &&&&&n&&&&&\\
 & 1 & 2 & 3 & 4 & 5 & 6 & 7 & 8 & 9 & 10 \\ 
\hline\hline
2 & 0 & 1 & 0 & 0 & 0 & 0 & 0 & 0 & 0 & 0 \\
3 & -1& 0 & 1 & 0 & 0 & 0 & 0 & 0 & 0 & 0 \\
4 & 0 & -2 & 0 & 1 & 0 & 0 & 0 & 0 & 0 & 0 \\
5 & 1 & 0 & -3 & 0 & 1 & 0 & 0 & 0 & 0 & 0 \\
6 & 0 & 3 & 0 & -4 & 0 & 1 & 0 & 0 & 0 & 0 \\
7 & -1 & 0 & 6 & 0 & -5 & 0 & 1 & 0 & 0 & 0 \\
8 & 0 & -4 & 0 & 10 & 0 & -6 & 0 & 1 & 0 & 0 \\
9 & 1 & 0 & -10 & 0 & 15 & 0 & -7 & 0 & 1 & 0 \\
10 & 0 & 5 & 0 & -20 & 0 & 21 & 0 & -8 & 0 & 1 \\
\hline
\end{tabular}
\caption{Some coefficients $a^{q}_{n}$ of the polynomial $\overline{U}_S(\bar{\mu},q)$ for the (S) single-defect case in the modulated $\mu_s$.}
\label{table:US-ChP}
\end{table}

\subsubsection{Commensurate and incommensurate potential}

When these two spatial distributions are applied to $\mu_s$, the function $U$ can be written as in Eq. (\ref{U_S}), where $\overline{U}_\lambda$ are now polynomials of degree $q$,  nonlinear in $\lambda$ and $\bar{\mu}$ that are still independent of $\Delta$. So far, we have not found recurrence formulas for them, but we can give examples for $q=3$. For the (C) commensurate case, we have 
\begin{equation}
 \overline{U}_C(\bar{\mu},q=3,\lambda) = \frac{1}{4}\lambda^2 (\lambda -3)\,\bar{\mu}^3 ,
\end{equation}

\noindent
while for the (I) incommensurate case, we have
\begin{eqnarray}
&\overline{U}_{I}(\bar{\mu},q=3,\lambda)   = & \nonumber \\
& - \left[ \cos(2\pi\beta) + \cos(4\pi\beta) + \cos(6\pi\beta) \right] \lambda (\bar{\mu}-\bar{\mu}^3 ) & \nonumber \\
& + \left[ \cos(2\pi\beta)\cos(4\pi\beta) + \cos(2\pi\beta)\cos(6\pi\beta) \right. & \nonumber \\
& + \left. \cos(4\pi\beta)\cos(6\pi\beta) \right] \lambda^2 \bar{\mu}^3  \nonumber & \\ 
& + \left[ \cos(2\pi\beta)\cos(4\pi\beta) \cos(6\pi\beta) \right]\, \lambda^3 \bar{\mu}^3 , & 
\end{eqnarray}

\noindent
where we have decided to keep the harmonics $\cos (2n\pi\beta)$ in the latter incommensurate case to show the structure of such polynomials.

\section{Modulated hopping amplitude}

The situation is very different when spatial distributions are applied to the hopping amplitude. Functions $U$ and $\Lambda_{\kappa}$ depend now on the disorder strength $\lambda$ and on the kind of distribution applied. Only the single-defect case has a factorization similar to the previous examples.

\subsubsection{Single-defect case}

When the single-defect spatial distribution is applied to the hopping, the function $U$ is factorized as in (\ref{tildeU_s}), where $\tilde{U}_{S}=\tilde{U}_{S}(\bar{\mu},q)$ is a polynomial function of degree $q$, whose coefficients are independent of  $\lambda$ and $\Delta$. There is no inhomogeneous effect for $q=2$ in this case. This polynomial $\tilde{U}_{S}$ has the same recurrence formula as in Eq. (\ref{eq:UH}), except that the largest-power non-zero coefficients are now $a^q_{q-2}=-1$ ({\it viz}.: $a^q_q=a^q_{q-1}=0$ for all $q$). Some of these coefficients are given in Table \ref{table:US-Hopp}. Interestingly enough, we  see that by shifting some rows and columns we find similarities between Tables \ref{table:US-ChP} and \ref{table:US-Hopp}.

\begin{table}[t]
\centering
\begin{tabular}{|c|ccccccccc|}
\hline
 q &&&&n&&&&&\\
 & 1 & 2 & 3 & 4 & 5 & 6 & 7 & 8 & 10 \\ 
\hline\hline
3 & -1& 0 & 0 & 0 & 0 & 0 & 0 & 0 & 0 \\
4 & 0 & -1 & 0 &  0& 0 & 0 & 0 & 0 & 0 \\
5 & 2 & 0 & -1 & 0 &0  & 0 & 0 & 0 & 0 \\
6 & 0 & 3 & 0 & -1 & 0 & 0 & 0 & 0 & 0 \\
7 & -3 & 0 & 4 & 0 & -1 & 0 & 0 & 0 & 0 \\
8 & 0 & -6 & 0 & 5 & 0 & -1 & 0 & 0 & 0 \\
9 & 4 & 0 & -10 & 0 & 6 & 0 & -1 & 0 & 0 \\
10 & 0 & 10 & 0 & -15 & 0 & 7 & 0 & -1 & 0 \\
11 & -5 & 0 & 20 & 0 & -21 & 0 & 8 & 0 & -1 \\ 
\hline
\end{tabular}
\caption{Some coefficients $a^q_n$ of the polynomial $\tilde{U}_{S}(\bar{\mu},q)$ for the (S) single-defect case in the modulated $t_s$.}
\label{table:US-Hopp}
\end{table}

The sum rule of the $\Lambda_{\kappa}$ functions in this case is now $\lambda$-dependent, as given in Eq.\ (\ref{Lambda}). The selected polynomial $\Lambda_0(\Delta,q,\lambda)=\sum_{n=0}^{q} b_n^q \,\Delta^n$ is again an even polynomial in $\Delta$, whose coefficients are however nonlinear in $\lambda$. Some of these coefficients can be seen from Table \ref{table:LambdaS-Hopp}. We notice that when $\lambda=0$ they match those in Table \ref{table:LambdaHomo}.

\begin{table}[ht]
\centering
\begin{tabular}{|c|ccc|}
\hline
 q &&n&\\
 & 0 & 2 & 4  \\ 
\hline\hline
2 & \ $(2+\lambda)^2$ \ & 0 & 0   \\
3 & \ $2(1+\lambda)$ \ & $2(3+\lambda)$ & 0    \\
4 & \ $-\lambda^2$ \ & $(4+\lambda)^2$ & 0    \\
5 & \ $2(1+\lambda)$ \ & $4(5+3\lambda)$ & $2(5+\lambda)$   \\
6 & \ $(2+\lambda)^2$ \ & $2(12+8\lambda-\lambda^2)$ & $(6+\lambda)^2$   \\
7 & \ $2(1+\lambda)$ \ & $6(7+5\lambda)$ & $10(7+3\lambda)$   \\
8 & \ $-\lambda^2$ \ & $(64+48\lambda+3\lambda^2)$ & $ \ (128+64\lambda-3\lambda^2)$ \   \\
\hline
\end{tabular}
\caption{Some coefficients $-b_n^q$ of the function $\Lambda_{0}(\Delta,q,\lambda)$ for the (S) single-defect case in the modulated $t_s$.}
\label{table:LambdaS-Hopp}
\end{table}

\subsubsection{Commensurate and incommensurate potentials}

Different from the above descriptions, the function $U$ cannot be factorized in these two latter cases. Therefore, the modulation effect can be simply written as in Eq.\ (\ref{tildeU_c}), where we have $\tilde{U}_C(\bar{\mu},\Delta,q,\lambda)$ and $\tilde{U}_I(\bar{\mu},\Delta,q,\lambda)$, respectively. There are no simple recurrence formulas found for these  cases. Yet, an example for $q=5$, for the (C) commensurate case, is given by
\begin{eqnarray}
& \tilde{U}_{C}(\mu,\Delta,q=5,\lambda) =  & \,\,\,\,\,\, \nonumber \\
& - 5\Delta^2 \lambda^2 \mu + 5\frac{\sqrt{5}+7}{32} \lambda^4 \mu 
  -\frac{5}{2}\lambda^2 \mu^3 -5\frac{\sqrt{5}+1}{2} \lambda^2 \mu ,  & 
\end{eqnarray}

\noindent
while for the new function $\tilde{\Lambda}_0^C=\Lambda_{0}+ \Lambda_{0}^{C}$, we have 
\begin{equation}
 \Lambda_{0}^{C}(\Delta,q=5,\lambda) =  \frac{1}{8} \lambda ^2 
 \left(20 - 5\lambda^2 -\lambda^3 +60\Delta^2 \right),
\end{equation}

\noindent
where $\Lambda_{0}$ corresponds to the homogeneous case, as given in Table \ref{table:LambdaHomo}. Similarly, an example for $q=3$, for the (I) incommensurate case, is given by
\begin{eqnarray}
& \tilde{U}_{I}(\mu,\Delta,q=3,\lambda) = & \,\,\,\,\,\, \nonumber \\
& - \left[ 2\lambda\left( \cos(2\pi\beta) +\cos(4\pi\beta)+\cos(6\pi\beta) \right)\right. & \nonumber \\
& + \lambda^2 \left.\left( \cos^2(2\pi\beta) + \cos^2(4\pi\beta) + \cos^2(6\pi\beta) \right) \right] \mu \, , &
\end{eqnarray}

\noindent
while for the new function $\tilde{\Lambda}_0^I = \Lambda_{0} + \Lambda_{0}^{I}$, we have 
\begin{eqnarray}
& \Lambda_{0}^{I}(\Delta,q=3,\lambda) =& \nonumber \\
& -2\lambda (1+ \Delta^2) \left[\cos(2\pi\beta)+\cos(4\pi\beta) + \cos(6\pi\beta) \right] & \nonumber \\
& -2\lambda^2 \left[ \cos(2\pi\beta)\cos(4\pi\beta) + \cos(2\pi\beta)\cos(6\pi\beta) \right. & \nonumber \\
& + \left. \cos(4\pi\beta)\cos(6\pi\beta) \right] &  \nonumber \\ 
& -2\lambda^3 \left[\cos(2\pi\beta)\cos(4\pi\beta)\cos(6\pi\beta)\right] . &
\end{eqnarray}

\paragraph*{\bf Acknowledgments.--}
We would like to thank Drs.\ Germ\'an Sierra, J. Carlos Egues, Rafael Barfknecht, Ra\'ul Santos, and Sergio Valencia for enlightening discussions. This work was financed by the Brazilian agencies CNPq and CAPES and by the ``Centro de Excelencia Severo Ochoa'' Programme under grant SEV-2012-024 (Spain).

\bibliographystyle{apsrev4-1}
\bibliography{pra_perez}

\end{document}